**From Biological Precursors to Artificial Cognition:
Consciousness, Embodiment, and the MEM Architecture**

Janusz A. Starzyk[1], Wiesław L. Galus[2]

**Abstract:** This article asks under what conditions artificial intelligence could warrant a rational attribution of consciousness. Linguistic ability, multimodality, memory, planning, action control, and humanoid embodiment are not sufficient evidence of phenomenal experience. Biological precursors such as excitability, homeostasis, neural networks, and hierarchical representation instead identify functions whose counterparts may be engineered. The paper compares conventional LLMs, hybrid h-LLMs, vision-language-action systems, and embodied agents with the Motivated Emotional Mind (MEM) architecture. MEM adds receptor-grounded representations, regulatory self-monitoring and interoception, affect, semblion-based associative memory, and re-entry into lower sensory and interoceptive maps. A related formulation defines a phenomenal state as a dynamic, integrated state of the whole embodied system in which sensor-grounded modal content is recurrently stabilized and coupled to a suprathreshold interoceptive representation of the system's regulatory state. The mathematical model of MEM published on arXiv specifies implementable relations among these components and makes the proposal more precise and testable. Comparative and causal tests in humans, invertebrates, and artificial systems could strengthen or weaken the MEM capabilities. MEM is therefore presented as a falsifiable research program requiring empirical validation and ethical caution.

## 1. Aim and Scope of the Article

Machine consciousness is often framed as a binary question: either an artificial system is conscious or it is not. Biological evolution suggests a different starting point. Perception, regulation, memory, valuation, and action control emerged and became progressively integrated in increasingly complex organisms. This functional gradation, however, should not be confused with a gradation of phenomenality. Excitability, embodiment, neural networks, and hierarchical organization support increasingly complex information processing, but none of these properties alone establishes conscious experience. The central question is therefore which additional organizational and dynamic properties may be relevant to phenomenality.

Three issues must be kept distinct: mechanisms supporting adaptive behavior, properties associated with conscious access and cognitive control, and processes proposed to constitute subjective experience. Explaining adaptive behavior does not by itself explain conscious access or phenomenality, and functional similarity between biological and artificial systems does not establish phenomenal similarity.

The aim of this article is to develop an architectural framework that traces a possible path from biologically evolved organizational properties to candidate conditions of phenomenality and then to their implementation in artificial cognitive systems. Rather than assuming a linear evolutionary progression toward consciousness, we identify properties that became progressively integrated during biological evolution and examine how they may be reproduced, approximated, or combined through different engineering solutions. We then use these properties to derive architectural criteria within the Motivated Emotional Mind (MEM) model and apply them to contemporary and prospective artificial systems, with particular attention to embodied agents and cognitive humanoid robot.

The analysis considers conventional large language models (LLMs), hybrid LLMs (h-LLMs) coupled to components such as memory, world models, perception, planning, or control, vision-language-action (VLA) systems linking multimodal representations to action (Driess et al., 2023; Zitkovich et al., 2023; Kim et al., 2025), and embodied agents organized according to MEM. These classes do not constitute a mandatory technological sequence. VLA systems need not evolve from h-LLMs, and MEM-based agents need not derive from language models. The comparison instead asks which relevant organizational properties are already implemented in each class and what additional integration would be required for autonomous, embodied cognition.

MEM is not conceived as an additional module appended to an LLM or VLA system. It is a hypothesis concerning the organization of the cognitive system as a whole. Its central proposal is that perception, internal regulation, memory, valuation, learning, motivation, and action must be dynamically coupled. The architecture incorporates regulatory variables, interoception, affect, motivated learning, hierarchical associative memory, semblion formation, detection of procedural gaps, simulation of action outcomes, and top-down sensory reactivation supporting secondary perception (Starzyk, 2008,

[1] jstarzyk@wsiz.edu.pl, University of Information Technology and Management in Rzeszów, Department of Information Systems Applications, https://wsiz.edu.pl/kadra-akademicka/jstarzyk/
[2] wieslaw25@gmail.com, independent researcher

2010; Galus, 2026a; Galus & Starzyk, 2026). Humanoid morphology is not a requirement of MEM, but it provides a useful test case because it requires the integration of visual, auditory, tactile, and proprioceptive information with internal-state monitoring, motor planning, and the consequences of action (Haikonen, 2012).

The article does not attribute consciousness to any existing artificial system. Linguistic competence, multimodal processing, memory, action, or embodiment alone do not establish subjective experience. Such properties may nevertheless provide measurable indicators of increasingly integrated cognitive organization (Chalmers, 2023; Butlin et al., 2023; Farisco et al., 2024). Assessment should therefore concern the causal organization of processes within the complete architecture, avoiding both anthropomorphic attribution and a priori exclusion of artificial consciousness.

MEM is consequently presented as an architectural hypothesis rather than an established theory of consciousness. It proposes receptor-grounded representations, re-entry into sensory and interoceptive maps, and coupling of representational selection with regulatory state and affect as candidate conditions of phenomenality (James, 1884, 1890; Lange & James, 1967; Salice, 2022). Their necessity and sufficiency remain to be established. The distinctive MEM claim is not that any single component produces consciousness, but that cognition, needs, internal state, and action must become dynamically integrated within the whole system for the candidate conditions of phenomenality to arise.

The article therefore follows a specific scientific argument: evolutionary organization provides the biological basis for identifying relevant organizational properties; MEM integrates these properties into a unified architectural hypothesis; the hypothesis is translated into explicit architectural criteria; these criteria can be implemented in artificial systems; their consequences generate experimentally testable predictions; and the resulting predictions provide opportunities for confirmation, comparison with alternative explanations, and falsification.

The analysis addresses three central questions: (1) which biological mechanisms should be regarded as precursors of cognitive complexity and which may be relevant to MEM-based phenomenality; (2) to what extent contemporary AI architectures implement these organizational properties; and (3) what additional organization would an artificial agent require to satisfy the candidate MEM criteria, particularly those involving embodiment, internal regulation, motivated cognition, re-entry, and phenomenality.

## 2. From Biological Precursors to Integrated Cognitive Organization

The biological evolution of consciousness need not be understood as the emergence of a single isolated trait, but rather as the gradual integration of mechanisms enabling an organism to interact with its environment in increasingly complex ways, regulate its own state, learn, anticipate the consequences of action, and integrate information about the world with information about itself. From this perspective, biological precursors of consciousness and corresponding organizational thresholds can be distinguished, and these may also be implemented in artificial systems. These categories should not, however, be conflated. A precursor establishes conditions for a later organization; a threshold marks the emergence of a new functional relationship among components of the system; and a candidate indicator of phenomenality concerns a property that may be more directly associated with subjective experience. Yet no single precursor—excitability, embodiment, learning, centralization, memory, or recurrence—is treated in this article as a sufficient indicator of phenomenality.

### 2.1. Excitability and Coupling with the Environment

The most fundamental precursor is excitability, which enables unicellular organisms to respond to chemical gradients, light, temperature, and mechanical contact. It establishes the basic system-environment relationship: a change in the environment can induce a change in the organism's state and behavior. This is only a functional point of departure. Chemotaxis or phototaxis may be highly effective and adaptive, although their description requires neither a globally accessible representation, episodic memory, nor recurrent reconstruction of sensory content. A response to a stimulus alone does not justify attributing subjective experience to an organism.

Analogously, in artificial systems a sensory interface or the ability to respond to input does not by itself constitute a threshold of consciousness. It is, however, a condition for the emergence of more complex perception-action loops.

### 2.2. Embodiment, Integrity, and Homeostasis/Allostasis

The next step enabling subjective feeling is biological embodiment, understood as the organization of many parts into a single, relatively autonomous and regulated organism. As an organism becomes more complex, it must communicate among distant parts of the body, synchronize responses, control resources, and maintain internal parameters within ranges compatible with organismic integrity.

In this sense, homeostasis and allostasis are important evolutionary roots of valuation and affect. Cannon (1932) described homeostasis as the maintenance of the relative constancy of a system's internal environment, while Sterling and Eyer (1988) introduced allostasis as stability achieved through change. Allostasis requires a model of changing sensory conditions within the organism—a process called interoception. Sennesh et al. (2022) describe how interoception can provide feedback on the operation of allostatic mechanisms. Maimon et al. (2025) discuss what it means for a signal to be interoceptive in an artificial system and the significance of internal signals for self-regulation. Damasio (1994, 1999, 2018) and Solms (2021) developed the thesis that feelings are inseparably linked to regulation of the organism's state, while Panksepp (1998) identified deep affective systems as an evolutionarily older core of mental life.

Regulation introduces a normative perspective in the biological sense: some states sustain the organism's integrity, whereas others threaten it. Such normativity provides a basis for evaluating events as beneficial or harmful to the system. This does not mean, however, that every regulatory system has experiences. A thermostat can maintain temperature, and a simple controller can minimize deviation from a set point, without thereby becoming a subject of experience.

In artificial systems, the analogous threshold is not merely the possession of actuators or a 'body,' but the existence of an internal state whose changes causally affect the agent's perception, learning, action selection, and behavior.

### 2.3. Neural Networks, Learning, and Memory

A further precursor is a neural network capable of consolidating relationships among configurations of stimuli, actions, and their consequences. Associative learning, memory, and prediction increase behavioral flexibility because they allow future responses to be modified on the basis of prior experience.

Distributed patterns emerge in the network that make it possible to distinguish beneficial from harmful configurations and to modify future behavior. Predictive capacity should be understood here operationally, without assuming an explicit world model or an experienced expectation. This level is likewise not yet sufficient for consciousness: memory, learning, and prediction may occur in systems to which we do not attribute phenomenal experience.

The analogous threshold in an artificial system is not the number of parameters or the mere capacity to learn, but the emergence of persistent representations that modify the system's future processing and action as a function of its history.

### 2.4. Progressive Integration Across Evolutionary Lineages

Evolutionary comparison should not be presented as a unidirectional ladder leading from 'lower' organisms to humans, but as a mosaic process in which different organizational components relevant to the emergence of more complex forms of consciousness could arise, develop, and become integrated independently in different evolutionary lineages. Centralization of the nervous system, hierarchical organization of sensory processing, the capacity to learn complex contingencies, integration of interoceptive signals, and the development of extensive feedback connections therefore need not form a single universal phylogenetic sequence.

The important task is to determine which elements of biological organization may be treated as successive precursors or functional thresholds leading to an architecture capable of integrating perception, organismic state, memory, valuation, and action.

For each group analyzed, it is therefore necessary to establish separately: (a) the presence of receptors and regulatory circuits; (b) the capacity for associative learning, persistent behavioral modification, and motivational trade-offs; (c) the presence of anatomical and physiological bases of recurrence, especially content-specific reactivation of earlier representations; and (d) the conclusions that may be drawn only after adopting particular assumptions about the relationship between functional organization and consciousness.

In the simplest organisms, the capacity to respond to stimuli and maintain homeostasis predates the emergence of nervous systems. At this level, chemical, mechanical, and light-sensitive receptors may already couple the state of the environment to regulation of the behavior of a cell or organism.

Cnidarians possess distributed nerve nets that coordinate the entire body without a distinct centralized brain. The demonstration of classical conditioning and associative memory in the sea anemone Nematostella vectensis shows that simple learning does not require a centralized nervous system (Botton-Amiot et al., 2023). This represents an important step from excitability alone to plastic network coordination, in which prior interactions with stimuli can influence an organism's future responses. It demonstrates neither content-specific re-entry into ordered sensory maps nor the existence of phenomenal feeling.

In bilaterally symmetrical invertebrates, such as nematodes, annelids, and gastropods, processing becomes more clearly centralized in ganglia and longitudinal nerve tracts, while specialized receptors become more tightly coupled to movement selection, behavioral regulation, and memory. Long-lasting, site-specific sensitization in Aplysia and avoidance learning in earthworms indicate that the effects of earlier stimuli can persistently modify the organism's subsequent actions (Walters, 1987; Wilson et al., 2014). The existence of memory, regulation, and feedback loops must nevertheless be distinguished from experienced aversion: feedback loops in a small ganglionic system do not yet constitute evidence of re-entry in the MEM sense. The crucial demonstration would be that feedback reconstructs specific sensory content and links it to the organism's felt state.

In arthropods, centralization of the nervous system coexists with specialized structures for learning, navigation, and action selection. Mushroom bodies and other circuits integrating sensory information with memory and stimulus value are of particular interest. In Drosophila, a recurrent architecture regulating learning and feedback connections linking innate and acquired valence with short- and long-term memory have been described (Eschbach et al., 2020; Huang et al., 2024). Moreover, reward-dependent trade-offs in responses to potentially harmful thermal stimuli have been demonstrated in bumblebees (Gibbons et al., 2022). Such behaviors are especially interesting from the MEM perspective because they indicate a functional integration of stimulus, memory of previous experience, valence, and action selection.

Cephalopods represent an independent lineage in which nervous-system complexity increased substantially. They are characterized by large numbers of neurons, specialized brain lobes, highly developed sensorimotor integration, and partially distributed control of the arms. Conditioned place aversion, preference for a place associated with relief, and targeted protective behavior following injury provide stronger, multicomponent indicators of a pain-like state than withdrawal reflexes alone (Crook, 2021; Kuo et al., 2022). From the MEM perspective, cephalopods offer a particularly important test of convergence: if similar relationships among receptor maps, internal state, memory, recurrence, and flexible action can arise in an architecture different from the vertebrate brain, this weakens the assumption that a particular cortical plan, such as the mammalian one, is necessary. It instead strengthens the hypothesis that what is decisive is functional-biophysical organization and the manner in which information is integrated, rather than a specific anatomical implementation.

In vertebrates, further increases in complexity involved the elaboration of centralized sensory systems, multilevel representations, memory, action-selection mechanisms, and interoceptive and neuromodulatory signals. The most comprehensive evidence for hierarchical maps and feedback connections comes from the mammalian brain, where numerous feedback projections modulate earlier stages of sensory processing alongside the feedforward stream (Felleman & Van Essen, 1991; Lamme & Roelfsema, 2000). This does not, however, justify treating the mammalian cortex as the only possible implementation of consciousness.

The concept of unlimited associative learning (UAL) provides a useful but competing way of defining an evolutionary transition threshold. UAL encompasses open-ended learning of relationships involving novel or complex stimuli, trace and higher-order conditioning, and flexible binding of representations to value (Birch et al., 2020). UAL identifies a candidate behavioral-cognitive capacity, whereas MEM asks the additional question of how mechanisms are organized to link receptor-based content with interoception, affect, semblions, and content-specific re-entry.

The two approaches may therefore provide complementary rather than interchangeable axes of comparison. UAL may indicate the level of learning flexibility and associative integration, whereas MEM makes it possible to analyze whether organizational foundations exist for a more complex, integrated representation of the relationship among a stimulus, the organism, and its action.

Table 1 shows progressive integration across evolutionary lineages.

**Table 1. Increasing Functional Integration.**

| Group/organization | (a) Receptors, regulation, and coupling with the organism | (b) Learning, memory, and motivational trade-offs | (c) Recurrence and integration of representations |
|---|---|---|---|
| Unicellular organisms and systems without neurons | Chemical, light, and mechanical sensors; cellular regulation and homeostasis. | Adaptation, habituation, or taxis; no basis for complex inferences about experience. | No neural network or receptor maps in the MEM sense. |
| Cnidarians | Specialized receptors and a distributed nerve net coordinating bodily activity. | Associative conditioning and memory demonstrated in Nematostella; evidence of motivational trade-offs is limited. | Bidirectional connections are possible, but there is no evidence of content-specific re-entry into ordered sensory maps. |
| Nematodes, annelids, and gastropods | Central ganglia, longitudinal tracts, and specialized sensorimotor systems. | Sensitization, conditioning, and avoidance; motivational trade-offs require separate tests. | Feedback loops exist, but their content-related function and relation to sensory representations remain poorly established. |
| Insects and other arthropods | A centralized brain, mushroom bodies, and the central complex, together with neuromodulatory signals associated with organismic state. | Multimodal learning, state-dependent choice, and motivational trade-offs; protective behavior in some taxa. | Recurrent memory and valence circuits documented in Drosophila; receptor-level re-entry remains undetermined. |
| Cephalopods | An elaborate brain and partially autonomous peripheral arm systems; rich visual-tactile-motor integration. | Learning, memory, flexible problem solving, place aversion, and relief seeking. | Numerous feedback connections are likely; tests of the direction and content of reactivation are needed. |
| Vertebrates | Hierarchical sensory and interoceptive systems; extensive neuromodulatory regulation. | From simple conditioning to planning and complex trade-offs; the range depends on the species. | Extensive feedback projections; content-specific re-entry is best studied in mammals. |

## 3. MEM Organizational Principles and Candidate Thresholds of Phenomenality

### 3.1. Hierarchical Representation and Information Integration

The next level of organization is a layered representational structure that enables progression from local features to more abstract categories and schemas (Felleman & Van Essen, 1991; Young, 2002). Deep artificial networks implement an analogous functional abstraction, while systems using motivated learning can learn to represent, generate, select, and solve their own problems (Starzyk & Raif, 2011; Starzyk & Graham, 2015). Motivated learning in autonomous systems is not yet an argument for their consciousness (Starzyk & Raif, 2010, 2011; Galus, 2025a, 2025b; Galus & Starzyk, 2020). Hierarchy supports generalization because many different episodes can be mapped onto a common type representation.

Information compression and abstraction are multipurpose mechanisms: they support recognition, prediction, and control even in systems to which we do not attribute experience. An important step beyond hierarchical representation is therefore to link representations of the world with representations of the system's own state and with possible consequences of action.

### 3.2. Interoception, Affect, Valuation, and Regulation

MEM's first candidate threshold of special relevance to consciousness couples world representations with interoception, affect, and a regulatory self-model. A stimulus then acquires state-dependent meaning: hunger, pain, risk, or anticipated benefit (Damasio, 1994; Critchley et al., 2004; Namkung et al., 2018; Graham et al., 2014, 2015; Starzyk et al., 2017). Internal signals motivate action and learning (Starzyk, 2008, 2010), while intrinsically motivated agents show how such principles may be implemented (Colas et al., 2022). Unlike an external reward alone, integrity-dependent valuation can change attention, memory, exploration, and choice with the agent's current state. In MEM, affect is this dynamic regulatory modulation of processing.

### 3.3. Semblions: Integrated Representations of Perception, Valuation, and Action

The next threshold is the concept of semblions: dynamic associative structures integrating a perceptual trace, its generalization, cross-modal relationships, valence, motivational context, and possible actions. Their organization is hierarchical and heterarchical, combining levels of categorization with lateral links among distant representations (Galus & Starzyk, 2020, 2026). A semblion is neither a single cell nor a static symbol: its context-sensitive activation connects perception with prior experience, internal state, valuation, and action.

### 3.4. Re-entry and Secondary Perception

The next postulated threshold is re-entry. Recurrent Processing Theory associates consciousness with feedback that modifies sensory activation (Lamme & Roelfsema, 2000; Lamme, 2006, 2018, 2020). MEM makes the more specific proposal that re-entry into lower sensory and interoceptive maps is necessary for secondary perception and may be necessary for phenomenality (Galus & Starzyk, 2020; Galus, 2022, 2023a, 2023b). A selected higher representation can feed back to reconstruct lower-level patterns, re-embedding an interpretation in sensory and interoceptive maps.

Overlaps between perception and imagery support this hypothesis without proving it (Albers et al., 2013; Dentico et al., 2014; Dijkstra et al., 2017, 2020; Naselaris et al., 2015). Tests must distinguish generic recurrence from feedback-driven reinstatement by measuring the location, direction, and modality-specific content of information flow.

Re-entry matters in MEM only as part of an integrated architecture: it regrounds selected representations in sensory and interoceptive processes already linked to valuation, memory, and action.

### 3.5. Language, Narrative, and Symbolic Reconstruction

Language, narrative, and symbolic logic extend cognition but are not conditions of phenomenal consciousness. MEM treats embodied schemas as higher-order semblions formed from sensorimotor and interoceptive experience (Harnad, 1990; Johnson, 1987; Lakoff & Johnson, 1980, 1999; Barsalou, 1999, 2008; Galus, 2024).

Linguistic signs can compress and recombine these structures into abstract concepts and plans, expanding report and reasoning without by themselves demonstrating feeling. Language is thus a later means of expressing and reconstructing experience, not its foundation. Rich symbolic processing may occur without interoception, regulation, embodiment, or re-entry.

### 3.6. From Biological Precursors to Candidate Organizational Thresholds

The preceding sequence is neither a rigid evolutionary order nor a construction recipe. Biology is genealogically continuous, but artificial counterparts may use different substrates and implementation paths.

It is therefore important to distinguish among a precursor, a functional organizational threshold, and a candidate condition of phenomenality. Excitability, embodiment, regulation, learning, memory, and hierarchical representation enable complex organization without warranting experience. MEM's candidate architecture arises only from their integration with interoception, valuation, self-state representation, semblions, and re-entry. Organizational thresholds may overlap, depend on one another, and arise in different orders.

Technological development may be continuous, but parameter count, computing power, and behavioral complexity do not automatically yield machine consciousness. The relevant comparison concerns feedback-integrated relations among perception, regulation, memory, valuation, self-state representation, and action.

Table 2 treats the integrated interoception-affect-semblion-re-entry organization as MEM's principal candidate threshold, while separating it from capacities sufficient only for complex cognition.

**Table 2. Comparison of System Classes with Respect to MEM Indicators**

| Biology | Analogous artificial organization | Status with respect to consciousness |
|---|---|---|
| Excitability | sensors + response | precursor |
| Embodiment | body + coupling with the environment | precursor |
| Homeostasis/allostasis | regulation of internal state | precursor |
| Neural network | processing network | precursor |
| Memory/learning | memory + adaptation | precursor |
| Hierarchical representations | hierarchical models | functional threshold |
| Interoception + affect | self-state model + valuation | candidate MEM threshold |
| Semblions | integrated representations | candidate MEM threshold |
| Re-entry | feedback to sensory/interoceptive maps | candidate mechanism of phenomenality |
| Language | LLM / symbolic layer | higher-level, nonessential layer |

MEM does not identify any single component with consciousness. It proposes that phenomenality becomes a scientifically testable hypothesis only when specific organizational properties become jointly integrated within an embodied, self-regulating cognitive system.

## 4. From Organizational Principles to Architectural Criteria

Disputes about LLM consciousness often arise from the ambiguity of terms. 'Understanding' may mean correct language use, generalization, a world model, effective action, or feeling. 'Consciousness' may mean access to information, reportability, phenomenal experience, self-consciousness, or moral status. Without these distinctions, linguistic competence can easily be mistaken for subjectivity, or a system can be denied any understanding simply because it lacks a biological body.

Block (1995) distinguished access consciousness from phenomenal consciousness. The former concerns information useful for reasoning, planning, reporting, and behavioral control; the latter concerns the subjective character of experience—what it is like to see a color, feel pain, hear a melody, or experience fear. With respect to AI, it is safer to speak of access functions than of a system's access consciousness. This terminology preserves the value of Block's distinction without assuming that a mechanism of selection and reporting creates a subject.

Intelligence, agency, autonomy, motivation, and feeling must also be distinguished. Intelligence means effective problem solving; agency, the selection of actions; autonomy, their sustained control without direct external direction; motivation, internal criteria of priority; and feeling, the phenomenal character of states. The distinction is empirically important: each property requires different tests and may be independently impaired.

Likewise, descriptive semantics is not constitutive semantics. A model may accurately describe pain because it knows relations among texts. The constitutive meaning of pain, by contrast, would require coupling the system's own damage signal to the protection of integrity, a change in priorities, and negative valence (Harnad, 1990; Barsalou, 1999). In an artificial system, it would therefore be necessary to show that violation of a protected variable not only generates a message but reorganizes processing, protective behavior, and subsequent learning.

We therefore adopt the four levels of description: (1) embodied agency, representations are coupled to perception and action; (2) functions of access, reporting, and control; (3) information processing; and (4) the postulated MEM architecture, comprising receptor and interoceptor grounding, affective modulation, and re-entrant reconstruction. Only the final level proposes a hypothesis about the mechanism of experience. These levels may partly overlap or diverge. A system may possess global reporting without embodied agency, whereas a simple organism may regulate its own state without language or elaborate reporting. Table 3 applies the same principle to classes of AI.

**Table 3. Comparison of Artificial-System Classes with Respect to MEM Indicators**

| System class | Access, reporting, and action | Self-regulation and re-entry | Permissible conclusion |
|---|---|---|---|
| Conventional LLM | Rich linguistic functions, reasoning, and reporting dependent on the input context. | No persistent needs, regulatory self-monitoring, or reconstruction of receptor maps. | Advanced information-access functions; no basis for attributing subject-level consciousness. |
| h-LLM / VLA | An LLM connected to memory, perception, tools, or an action policy; a VLA maps images and language onto actions. | May possess memory and perception-action coupling, but usually lacks its own regulatory pressure and re-entry in the MEM sense. | Embodied functional agency is not in itself an indicator of phenomenality. |
| Embodied robot | Continuous perception, proprioception, and motor control in a physical environment. | Telemetry may protect the system, but without coupling to priorities, learning, and memory it does not constitute technical interoception. | Neither multimodality alone nor humanoid morphology is sufficient to attribute experience. |
| MEM-like agent | Integrates perception, interoception, action, associative memory, and program simulation. | Postulated indicators: regulatory self-monitoring, affect as regulatory pressure, and re-entry into sensory and interoceptive maps. | More architectural requirements indicative of phenomenal consciousness; no evidence that they constitute a sufficient condition. |

Considering brain phylogeny, convergent architectures, and the development of cognitive theories, we may ask which minimal organizational relationships—developed in different combinations over the course of evolution—should be reproduced in an artificial-system architecture before it can be said to implement functional mechanisms postulated by MEM.

Table 4 synthesizes the properties of different cognitive architectures and identifies their limitations and capabilities.

**Table 4. Limitations and Capabilities of Cognitive Architectures**

| Mechanism | Biology | LLM | Hybrid AI | VLA | MEM agent |
|---|---|---|---|---|---|
| Receptor grounding | ✓ | limited | partial | ✓ | ✓ |
| Embodiment | ✓ | undetermined | partial | ✓ | ✓ |
| Interoception | ✓ | undetermined | possible | limited | central |
| Regulation of the system's own state | ✓ | undetermined | partial | partial | ✓ |
| Affect as regulatory value | ✓ | simulated | possible | possible | central |
| Associative memory | ✓ | partial | ✓ | ✓ | ✓ |
| Semblions | postulated | ? | ? | ? | ✓ |
| Re-entry into lower-level maps | ✓ | undetermined | possible | partial | central |
| Secondary perception | postulated | undetermined | undetermined | undetermined | candidate |
| Phenomenal evidence | unresolved | none | none | none | unresolved |

Philosophical problems such as the explanatory gap, mind-body identity, mental causation, and qualia require conceptual analysis joined to mechanisms. Functional or correlational descriptions alone do not explain subjective character (Nagel, 1974; Levine, 1983; Chalmers, 1995, 1996), but they permit measurement and comparison of models.

Major theories isolate different mechanisms. GNWT explains global availability across frontoparietal networks (Baars, 1988; Dehaene & Naccache, 2001; Dehaene & Changeux, 2011; Mashour et al., 2020); HOT invokes higher-order representation (Lau & Rosenthal, 2011); IIT formalizes integrated causal power (Tononi, 2004; Tononi & Koch, 2015; Albantakis et al., 2023); Predictive Processing and Active Inference emphasize hierarchical inference, prediction error, and action (Rao & Ballard, 1999; Friston, 2010; Clark, 2015; Hohwy, 2013); and RPT emphasizes sensory feedback (Lamme & Roelfsema, 2000; Lamme, 2006, 2018, 2020).

These theories illuminate access, metarepresentation, integration, prediction, or recurrence without fully identifying why a particular process is felt as color, pain, or emotion. GNWT is strongest on reportability and broad access; HOT on metarepresentation; IIT on formal integration, though mapping its quantities to particular qualities is difficult; PP on hierarchical construction; and RPT on feedback. Active Inference further relates uncertainty reduction to survival and regulation, including exploratory and exploitative behavior (Friston et al., 2016, 2017; Pezzulo et al., 2024). MEM's distinctive burden is to show that receptor-grounded, interoceptively coupled re-entry explains more than these existing mechanisms separately or in combination.

MEM integrates regulatory interoception, affective modulation, associative memory, and content-specific feedback, and hypothesizes that conscious recurrence reactivates lower sensory and interoceptive maps. To turn this claim into an architectural criterion, the phenomenal target itself must be defined more precisely. Table 5 specifies what exactly must be present in the system for us to say that it meets the MEM criteria.

**Table 5. System Properties According to MEM**

| MEM principle | Biological interpretation | Architectural requirement | Observable prediction |
|---|---|---|---|
| receptor grounding | sensory coupling | persistent sensorimotor loop | dependence on grounded input |
| interoception | internal regulation | internal state variables | state-dependent cognition |
| affect/valuation | significance | global modulation | altered processing according to internal needs |
| semblion | integrated representation | multimodal binding | persistent cross-modal representation |
| re-entry | recurrent sensory interaction | top-down/bottom-up loops | reactivation effects |
| secondary perception | reconstruction | internally generated perceptual state | perception without external stimulus |
| procedural gaps | unmet action requirements | goal/action monitoring | self-initiated learning |

A related MEM analysis published in Qeios defines the target state as follows: A phenomenal state is a dynamic, integrated biophysical state of the whole embodied system in which sensor-grounded modal content is recurrently stabilized and coupled with a suprathreshold interoceptive configuration representing the system's current regulatory state. The modal content may be elicited by an external stimulus, arise within the system, or be reconstructed top-down through recurrent processing (Galus, 2026b).

Phenomenal consciousness is correspondingly the capacity of an embodied system to instantiate such states; occurrent phenomenal consciousness consists in their temporally organized and continuously changing integration (Galus, 2026b). For machine assessment, the implication is decisive: sensory input, recurrence, or telemetry considered separately do not meet the definition. The modal and regulatory components must form one causally integrated, independently measurable whole-system configuration.

## 5. Mathematical Formulation of MEM Organizational Stages

The MEM formalism is not a complete algorithm of consciousness but an implementable description of receptor activation, representational selection, regulation, top-down reconstruction, and action-program formation. Its full equations, definitions, and notation are available in the arXiv paper 'A Mathematical Model of Motivated Emotional Mind: A Cognitive Embodied System' (Galus & Starzyk, 2026); only the equations required here are reproduced. By specifying variables, couplings, thresholds, and decision rules, the formalization constrains the verbal theory, enables simulation and ablation, and thereby strengthens the scientific credibility of the present proposal. It does not replace empirical validation.

The equations define the functional components of the architecture. An LLM can implement representational matching and competition without receptor grounding; a VLA can add perception and action without self-regulation; a MEM-like agent would combine these functions with self-monitoring and re-entry.

The first stage integrates sensory activation with lateral, feedback, and interoceptive influences. A typical LLM primarily performs complex transformations of internal representations; a robotic h-LLM may add receptor inputs, whereas a MEM architecture additionally requires that these inputs be linked to interoception, regulatory context, and re-entrant pathways.

The second stage is associative recognition. Semblion m stores a sensory prototype $\mu_m$, a bodily-motivational context $\rho_m$, and a valence $v_m$. The degree to which the current signal matches such a representation is defined by Equation (6):

$$u_m(t) = \alpha_s \mathrm{sim}\left(h_{\mathrm{sens}}^{ff}(t), \mu_m\right) + \beta_b \mathrm{sim}(b(t), \rho_m) + \xi_v v_m \qquad (6)$$

Unlike ordinary classification, the result $u_{\mathrm{m}}(t)$ depends not only on the similarity of the current feedforward representation $h_{\mathrm{sens}}^{ff}(t)$ to the prototype $\mu_{\mathrm{m}}$, but also on the correspondence between the current context $b(t)$ and the stored context $\rho_{\mathrm{m}}$ and on the valence $v_{\mathrm{m}}$. The same object may therefore receive a different priority under hunger, pain, threat, or curiosity. This constitutes an important organizational advance beyond a system that recognizes only statistical regularities in data.

The third stage is competition and representational selection. Selection determines which representations achieve functional dominance and can control subsequent processing, attention, and top-down reconstruction. In successive generations of machine systems, this corresponds to a transition from considering many possibilities in parallel to a stable choice of interpretation conditioned by context and motivation.

The fourth component is allostatic regulation. In the MEM model, needs are not replaced by an arbitrary reward, but arise when variables exceed their lower or upper tolerance limits $n_{\mathrm{r}}(t)$. The original one-sided measures of violations, global affect, and regulatory cost are defined, respectively, by Equations (2)–(4):

$$\Delta n_r^-(t) = \max\left(0, n_r^{\min}(t) - n_r(t)\right), \qquad (2)$$

$$\Delta n_r^+(t) = \max\left(0, n_r(t) - n_r^{\max}(t)\right), \quad r = 1, \ldots, R.$$

The complete vector of regulatory violations is:

$$d(t) = (\Delta n_1^-(t), \ldots, \Delta n_R^-(t), \Delta n_1^+(t), \ldots, \Delta n_R^+(t))^\top.$$

$$A(t) = g\left(d(t)\right) = \tanh(w_A^\top d(t) + c_A). \qquad (3)$$

$$C(t) = \sum_{r=1}^{R} [\alpha_r (\Delta n_r^-(t))^2 + \beta_r (\Delta n_r^+(t))^2]. \qquad (4)$$

In Equation (3), the violation vector $d(t)$ is transformed into a bounded global affective-regulatory signal $A(t)$, while $C(t)$ from Equation (4) assigns different weights to deviations below and above the tolerated ranges. Formally, $A(t)$ expresses the integrated regulatory pressure arising from violations of multiple variables relevant to the system's integrity. Its affective interpretation is inspired by the James–Lange perceptual theory of emotion, according to which emotional feeling is associated with the perception of changes in bodily state (James, 1884, 1890; Lange & James, 1967). At the functional level, $A(t)$ can therefore also be described as global affect, because it modulates learning, exploration, representational valence, action-program selection, and decision making.

The MEM model predicts that this signal can contribute to phenomenally experienced affect only when its interoceptive sources are represented in the appropriate maps and undergo re-entry and secondary perception. $A(t)$ denotes a global affective-regulatory context that gives representations and actions significance for the system itself, and thus constitutes affective regulatory pressure. The MEM interpretation of pain provides an example. Within MEM, the functional meaning of a pain counterpart for the system requires causal coupling of a damage signal to protection of integrity, changes in priorities, learning, and negative valence. This is an assumption of the model, not a generally accepted definition of pain or meaning.

The fifth stage, which is central to the MEM hypothesis of phenomenal consciousness, involves top-down reconstruction. Once a set of winning semblions has emerged, their prototypes reinstate a reconstructed sensory representation. MEM predicts that re-entry into sensory or interoceptive fields is a necessary condition of secondary perception—imagery, recall, or anticipation. We introduce its necessity for conscious experience as a new and significant hypothesis of the model.

The final stage concerns situations in which the system lacks a sufficiently effective action program. It then generates candidate programs and selects the program that maximizes value, equivalently minimizing predicted cost, as in Equation (31):

$$\Pi_{new}(t) = \arg\max_{\Pi \in \mathcal{P}_{cand}(t)} Q(\Pi \mid s_t) = \arg\min_{\Pi \in \mathcal{P}_{cand}(t)} J(\Pi \mid s_t). \qquad (31)$$

A new program is stored only if the increase in its decision value relative to the best program available in the current repertoire reaches or exceeds the improvement threshold $\theta_{imp}$. This threshold prevents the consolidation of programs whose apparent advantage may result from noise, uncertainty in the simulation, or random fluctuations in evaluation.

This sequence distinguishes a linguistically plausible plan from competence assessed by the consequences of action and regulatory cost. The equations do not prove consciousness; they isolate implementable mechanisms and couplings whose functional contribution can be assessed by selectively disabling individual architectural components in controlled ablation studies. The novelty of the MEM formalism does not lie in the mere use of similarity or expected cost, since these mechanisms are well known. It lies in coupling them within a single model to bodily-motivational context, re-entry, sensory reconstruction, regulatory affect, and consolidation of new programs. New organizational functions emerge along the path from symbolic processing to an embodied system with motivational selection, secondary perception, and the capacity to create new modes of action.

## 6. Where Current AI Architectures Stand

### 6.1. LLMs: Advanced Cognitive Functions and Architectural Limitations

Large language models are central to current discussions about machine consciousness. Transformers have enabled the learning of complex relationships in massive corpora, and LLM scaling has produced contextual generalization, tool use, and increasingly rich representations (Vaswani et al., 2017; Brown et al., 2020). They are not merely simple tables of word transitions. At the same time, linguistic competence, knowledge, and planning are cognitive functions; they do not constitute independent criteria of consciousness. In multimodal systems, the same mechanism can integrate text with images or sound, while in agentic systems it can invoke tools and execute successive steps of a plan. This expands the range of computationally accessible information, but does not by itself introduce a persistent center of goals or a regulatory perspective belonging to the system.

LLMs can categorize, construct analogies, explain, and perform multistep reasoning. Information is used broadly within the current computation, which partly resembles functions associated with access consciousness. This functional similarity does not, however, establish a unified subject of access. The input context may serve as short-term working memory, but its content is usually supplied externally, limited to a session, and lacks autonomous consolidation. Memory extensions alter this situation functionally, but still require a separate analysis of agency and subjectivity.

LLMs can also operate with image schemas in a linguistic sense because text preserves traces of human embodied experience (Lakoff & Johnson, 1980, 1999). A model may correctly use the CONTAINER or SOURCE-PATH-GOAL metaphor, but this does not imply that the schema was formed through its own movement, touch, balance, and regulation (Galus, 2024). This is an instance of inherited grounding: semantic relationships originate in descriptions of other subjects' experiences. They may support accurate reasoning about the world even though they were not acquired through the model's own perception-action loop or connected to its own regulatory cost.

Their semantics is therefore grounded largely indirectly in texts and data produced by embodied humans. The debate need not force a choice between ‘mere imitation’ and full human understanding: LLMs may possess rich functional semantics without their own receptor-affective semantics (Harnad, 1990; Bender et al., 2021; Shanahan, 2024). Functional semantics can be assessed through transfer, counterfactual reasoning, and robustness to changes in wording. Receptor-affective semantics would additionally require demonstrating that the meaning of a concept depends on the system's own history of perception, action, and regulation.

LLMs implement advanced functions of information access, reasoning, and reporting that are partly analogous to mechanisms associated with access consciousness, but this does not justify attributing consciousness to them in a subject-level sense. A conventional LLM lacks persistent goals of its own, autonomous control, and a unified subject of access; its operation is organized by the prompt, context, and generation mechanism (Block, 1995; Chalmers, 2023). A model's statement that it remembers, wants, or feels something is therefore primarily a generated report. Its evidential value would depend on independent confirmation of the architecture, persistence of the state, and causal role of the reported content.

From the perspective of the MEM hypothesis, current LLMs also fail to implement many candidate indicators of phenomenality: their own receptor fields, regulatory interoception, affect, and re-entry into sensory maps. An image as an input matrix and a textual description of pain are not yet the system's own perception, nociception, or feeling. A multimodal encoder may, of course, create representations shared across text and images. The critical MEM question, however, is whether a top-down representation can reconstruct a field associated with the system's own sensor and whether that reconstruction participates in the system's regulation and learning.

LLMs are therefore an important stage in the development of machine cognition and evidence for the partial multiple realizability of linguistic-narrative functions. They are not, however, a stage of consciousness in the subject-level sense, and their achievements do not settle the conditions of phenomenality (Galus, 2026a). This allows their accomplishments to be analyzed without disparaging machine intelligence or equating it with experience.

### 6.2. Hybrid LLMs: Persistent Memory, World Models, and Tool Use

In this paper, the term h-LLM is used as a working designation for a broad class of hybrid architectures in which an LLM constitutes one component of a larger cognitive or decision-making system. This includes, among others, agentic systems, language models equipped with extended memory mechanisms, robotic systems that use LLMs, architectures integrating world models, and systems capable of using external tools and resources.

### 6.3 VLA Systems: From Multimodal Processing to Embodied Agency

A partly related but more precisely defined direction is represented by VLA models: PaLM-E incorporates continuous sensory data into a multimodal model, RT-2 represents robot actions as tokens, and OpenVLA learns a vision-language-action policy from robotic demonstrations (Driess et al., 2023; Zitkovich et al., 2023; Kim et al., 2025). These approaches all extend a language model with data streams and an action space, although their technical organization differs. PaLM-E supports, among other functions, embodied reasoning, whereas RT-2 and OpenVLA learn a more direct mapping from observations and instructions to robot control.

Such systems can accumulate perception-action-outcome episodes and use them in planning. This makes it possible to study concept grounding, transfer across tasks, and plan correction on the basis of physical constraints without inferring experience from these capacities. Episodic memory, however, requires more than an archive of logs: an event should retain its temporal organization, the agent's state, the action performed, and its consequences. Only such a record makes it possible to distinguish repetition of an instruction from learning from one's own failure.

Connecting a model to a camera and a robotic arm provides an instrumental body. For the body to play a regulatory role, energy loss, overheating, or risk of damage must causally alter priorities, learning, and action selection. Battery or temperature telemetry alone is diagnostic information, not technical interoception or feeling. The causal-coupling requirement can be tested through interventions: after the self-monitoring channel is disconnected, the agent should systematically change its protective strategies, and after the signal is restored, it should recover them without the addition of a new external reward.

### 6.4 What Current Systems Still Lack Relative to MEM

h-LLMs and VLA systems may achieve substantial functional agency. Multimodality, recognition, and action control alone do not, however, provide a sufficient basis for attributing a phenomenal world to a robot. The MEM hypothesis requires additional regulatory and re-entrant couplings, and their presence would still constitute an indicator rather than proof. This distinction is especially important when interpreting achievements in robotics: execution of a complex instruction demonstrates integration of perception, language, and control, but does not yet show whether error, effort, or damage has a subjective character for the system.

The research value of these systems lies in testing whether they can compare a textual plan with its execution, learn concepts such as “slippery” through direct interaction, and connect language with schemas of force, path, balance, and obstacles. These are tests of grounded generalization, not of consciousness. For example, an agent may learn to increase grip force after detecting microslip and apply this rule to new objects. Such a result demonstrates embodied generalization; only further manipulations of regulation and re-entry can test the contribution of mechanisms specific to MEM.

Embodied cognition has long treated perception, movement, and the environment as coupled processes (Varela et al., 1991; Clark, 1997; Noë, 2004; Chemero, 2009; Gibson, 1979; Starzyk & Graham, 2015). MEM adds regulatory self-monitoring, affect, and secondary perception to this program. In these approaches, the body is an active component of computation because movement constraints and action possibilities jointly shape representations.

Humanoid morphology is not necessary. It facilitates operation in environments designed for humans and comparison of sensorimotor repertoires, but candidate MEM conditions can also be studied in wheeled robots, manipulators, or simulated agents. The more neutral term ‘embodied agent’ therefore captures the scope of the thesis better than “MEM-compliant humanoid”’

## 7. A MEM-Compliant Artificial Agent

A MEM agent would require stable ***exteroceptive*** *and* ***proprioceptive*** maps, perception and receptor-***grounded representation***, as well as self-monitoring channels that (a) track variables relevant to the system's integrity and resources, (b) are causally coupled to regulation, attention, learning, and ***action planning***, and (c) create a memorable context of consequences. A battery, temperature, or error reading without such coupling remains telemetry.

The system would also require hierarchical ***associative memory***. Semblions would link sensory, motor, procedural, linguistic, and affective traces. A semblion of 'slipperiness,' for example, would encompass surface appearance, microslips, changes in grip force, prior corrections, and the risk of losing an object, rather than merely a dictionary definition (Galus, 2023b).
Another proposed component is ***re-entry***: activation of a higher-order semblion should be able to reconstruct patterns in lower-level sensorimotor and interoceptive maps. MEM predicts that such reconstruction is necessary for secondary perception and imaginal evaluation of a plan; whether and when it would have a phenomenal character remains a hypothesis.

Procedural memory should distinguish a generalized program from a single episode. A program is a sequence of elementary transition operators $\mathcal{T}_{\mathrm{j}}$ defined in the mathematical MEM model as

$$\Pi = \left(\mathcal{T}_{j_0}, \mathcal{T}_{j_1}, \dots, \mathcal{T}_{j_{T-1}}\right). \tag{15}$$

Episodic memory, by contrast, preserves a temporally ordered trajectory of states and actions, represented by Equation (16):

$$E = \left(s_0, a_0, \mathcal{T}_{j_0}, s_1, a_1, \mathcal{T}_{j_1}, \dots, a_{T-1}, \mathcal{T}_{j_{T-1}}, s_T\right), \tag{16}$$

where $s_t = (h(t), b(t), n(t))$ denotes the system's complete representational, ***bodily-motivational, and allostatic state***, $a_t$ is the action performed, and $\mathcal{T}_{j_t}$ is the elementary transition operator leading to the next state. This distinction separates the memory of a specific event from generalized competence. An LLM can describe a sequence of steps, whereas a MEM-like agent should be able to execute or counterfactually simulate a program within its own state space and evaluate its value on the basis of predicted regulatory cost.

The system should also handle ***procedural gaps***: situations in which no available program reaches the required value. The mechanism then generates combinations of known operators, simulates their consequences, and consolidates a solution once the improvement criterion is satisfied.

As described in Section 5, global affect $A(t)$ denotes, in the MEM formalism, a bounded signal of motivating regulatory pressure that can modulate learning, exploration, consolidation, and representational priority. It is neither the valence of a single semblion, nor a measure of physiological arousal, nor feeling itself. MEM further predicts that coupling this pressure with re-entry and secondary interoceptive perception constitutes a hypothetical condition of affective phenomenality. Moreover, it is not sufficient merely to add separate modules for interoception, memory, emotion, and perception. Their causal and dynamic coupling is essential.

Satisfying these requirements would not automatically make an agent conscious or human-like. It would merely justify a more cautious indicator-based assessment of an alternative architecture. Different sensors, bodily dynamics, and learning histories might lead to different forms of organization whose subjective aspect would remain uncertain (Farisco et al., 2024).

Classical identity theory holds that an experience is identical to the corresponding brain process or state. MEM, however, shifts the focus to the organization of the ***entire dynamic system***. It is precisely this that later allows for a meaningful transition to the concept of multiple realizability.

## 8. Testing MEM: Predictions, Experiments, and Falsification

### 8.1. Epistemic Status and the Principle of Strong Inference

The proposed studies test the functional and architectural predictions of MEM, not consciousness directly. From a third-person perspective, one can measure behavior, network dynamics, the direction of information flow, the effects of interventions, and model fit, but not the subject's feeling itself. A result should therefore alter the degree of empirical justification for attributing phenomenality rather than establish a binary 'conscious–unconscious' verdict. It is especially important to distinguish indicators of phenomenality from cognitive functions that may occur without experience.

Adversarial collaboration, as applied in the direct comparison of GNWT and IIT, may serve as a methodological model: advocates of the theories and independent researchers jointly specified divergent predictions, success criteria, and the interpretation of results before data analysis, and then tested them using multiple methods and independent samples (Cogitate Consortium et al., 2025). An analogous program for MEM should preregister: (a) critical and auxiliary predictions; (b) the predicted direction, timing, and location of the effect; (c) the minimum effect regarded as informative; (d) a result that would weaken the model; and (e) a procedure for adjudicating mixed results. Data collection and the primary analysis should be conducted by teams not involved in formulating MEM, and the sample should be divided into an optimization subset and an independent confirmatory subset.

The hypotheses being compared are not entirely mutually exclusive. RPT predicts the importance of local recurrent processing in sensory cortex; predictive processing (PP) predicts top-down generation of expectations and the encoding of prediction errors; and GNWT/GWT predicts late, nonlinear ‘ignition’ of a representation and its global availability. MEM may coexist with each of these mechanisms. The mere detection of recurrence, prediction, or global broadcasting therefore cannot count as support specific to MEM. MEM's critical prediction concerns a more specific coupling among these processes: content-specific re-entry is to reinstate a representation in a receptor-grounded lower-level map, and the strength and consequences of this reconstruction are to depend on regulatory self-monitoring, global affect, associative memory, and action possibilities.

### 8.2. A Common Framework for Controls and Analysis

Human studies should combine threshold and suprathreshold conditions, immediate and delayed reports, and no-report conditions in which changes in the percept are estimated indirectly, for example through optokinetic responses, pupil diameter, or autonomic responses. This helps separate perceptual processes from decision, introspection, and the motor execution of a report, although no-report indicators also remain inferential (Frässle et al., 2014). High-temporal-resolution EEG or MEG should preferably be combined with laminar fMRI and, in clinically justified cases, with iEEG. Correlational evidence should be supplemented with safe causal intervention, for example by TMS, temporal masking, reversible pharmacological modulation, or manipulation of coupling between regions.

In studies of artificial systems, a complete MEM-like agent should be compared at least with a feedforward architecture, a recurrent world model (Ha & Schmidhuber, 2018), model-based reinforcement learning (Moerland et al., 2023), homeostatic reinforcement learning (Keramati & Gutkin, 2014), a predictive-coding/active-inference model, and a global-workspace variant. Comparisons should match data and the budgets for interaction and tuning, rather than only parameter count or operating speed. Parameter count, memory capacity, training data, the number of environmental interactions, access to sensors and actuators, and floating-point operations per episode must all be matched. The study should then sequentially disable functions characteristic of the MEM-like architecture, for example, interoceptive coupling, global affective modulation, re-entry, or episodic memory, while retaining the other architectural components. Such planned and selective disabling of one function is termed an ablation. To rule out the possibility that poorer performance following ablation results merely from reduced model capacity, a control variant with a comparable number of parameters and operations should be used. Parameters and computation freed by disabling the function under study may be allocated to additional layers or nonspecific memory that do not implement that function. This makes it possible to determine whether the observed effect depends on the specific functional organization postulated by MEM or only on greater computational capacity.

A critical question for assessing necessary conditions of consciousness will be whether selective ablation of a given function alters system behavior in the manner predicted by MEM while the remaining architectural functions and comparable learning conditions are preserved.

The basic unit of inference should therefore be not a simple main effect but a preregistered interaction among components. MEM predicts that receptor grounding, re-entry, and regulatory modulation produce an effect greater than the sum of their separate effects. The analysis should include mixed-effects models, confidence-interval estimation, causal mediation analysis, and Bayesian model comparison that includes hybrid variants. Learned systems require independent test sets, multiple random initializations, and reporting the distribution of results rather than the best run. A result consistent with MEM is diagnostic only when a simpler model with matched resources cannot explain it equally well.

### 8.3. Tests of Perceptual and Re-entrant Organization

**Test 1 Receptor-Grounded Re-entry**

The critical hypothesis is that consciously accessible sensory content will be associated not only with late recurrent activity, but also with content-congruent information flow from higher-order areas to lower-level maps that preserve receptor topography. In humans, backward masking, Kanizsa stimuli, partially occluded objects, and near-threshold stimuli can be used. Parameters should be adjusted individually so that the same physical class of stimulus is seen on some trials and unseen on others, with a comparable initial feedforward pass. Each trial should decode not only the object category but also map-based features—location, orientation, color, or spatial frequency—and determine the direction of information flow between higher-order and early sensory areas.

The findings of Fahrenfort, Scholte, and Lamme (2007) show that masking can attenuate late signals associated with re-entry, so such an experiment is simultaneously a test of RPT. Discrimination among the theories requires an additional manipulation: temporally precise TMS or another reversible perturbation should disrupt feedback to a specific lower-level map without comparably disrupting initial feedforward encoding, attention, or working-memory capacity. MEM predicts a selective reduction in content-specific reconstruction, subjective vividness, and subsequent use of that content in choices dependent on organismic state. RPT primarily predicts an association between visibility and local recurrence; GNWT predicts late availability of the content in the frontoparietal network; and PP predicts a change in predictive precision and prediction error.

Confirmation that visibility depends on feedback would support both MEM and RPT and, in part, PP. A threefold result would be more specific to MEM: (1) reinstatement of an exact feature in a lower-level receptor map, (2) the causal necessity of this reinstatement for secondary perception, and (3) modulation of its strength and consequences by the interoceptive regulatory state. If conscious vividness and flexible use of content remained unchanged after the effective elimination of re-entry defined in this manner, MEM's critical claim that it is necessary would be weakened.

**Test 2 Secondary Perception and Reactivation**

The second test should separate current receptor input from secondary perception. After associations have been learned, a neutral cue elicits imagery of a previously seen color and orientation, a previously heard tone, or a previously touched texture while the actual stimulus is absent. Decoders trained on real perception are used to determine whether the imagined content reactivates the same map and whether the direction of information flow is reversed relative to perception. Identification accuracy, vividness, delayed memory, and the effect of imagery on the next choice should be measured separately. Controls include a verbal-label-only condition, a semantic-memory condition without imagery, and disruption of maps in the relevant and irrelevant modalities.

MEM predicts that eliminating top-down reconstruction in a receptor map will reduce vividness and sensorimotor transfer even if the participant can still provide the verbal category. RPT may explain the need for recurrence, PP the generation of content by a top-down model, and GNWT the availability of imagery for reporting and multiple tasks. PP cannot, however, be equated with phenomenality because expectation-violation signals also occur in the primary visual cortex of anesthetized mice (Tang et al., 2023). MEM would receive additional specific support only if reactivation were receptor-topographic, causally necessary, and modulated by global affect and episodic history rather than merely by predictive accuracy.

### 8.4 Tests of Embodied Regulation and Significance

**Test 3 Interoception, Global Affect, and Significance**

In the third test, the external stimulus and its objective value remain constant, while internal state is varied within safe and reversible limits: satiety, respiratory load, temperature, fatigue, or the anticipated cost of effort. Global affect should be operationalized as global regulatory pressure calculated from multiple deviations, not as arousal, the valence of an individual stimulus, or experience itself. The study measures internal state, activity in interoceptive-affective networks, perceptual thresholds, preferences, exploration, learning, and memory persistence. The key question is whether the same image, sound, or action opportunity acquires a different weight and sensory representation depending on the organism's own state.

Control conditions include a neutral state; a change in arousal alone without regulatory consequences; an external reward of the same behavioral value; interoceptive information available for report but decoupled from decisions; and a sham manipulation. In an artificial agent, full regulatory coupling should be compared with mere battery or temperature telemetry that can be read but does not change priorities, learning, or the protection of integrity. Homeostatic RL and active inference should also be included because both model families predict behavior dependent on internal state.

The prediction specific to MEM is an interaction: interoceptive state changes global affect, modifies semblion competition and consolidation, and re-entry then changes content-specific reconstruction in a sensory or interoceptive map. Prediction-error reduction alone supports PP; a change in action policy alone supports homeostatic RL; and mere availability of state information may support GNWT. MEM is strengthened only by demonstrating the entire causal chain and selectively interrupting it. If disconnecting interoception from global affect had no effect while sensor quality and motor performance remained intact, the model's affective-regulatory core would be weakened.

**Test 4 Semblions, Episodic Binding, and Action Consequences**

The semblion test should assess not merely object recognition but the formation of a persistent structure linking multimodal features, context, internal state, and the action repertoire. The same object is learned in two contexts with opposite bodily or task-related consequences and is then presented in a new modality and after a delay. Representational similarity, episodic-trace reactivation, generalization to new exemplars, action selection, and changes in valence are measured. Ablations successively disconnect lateral connections, episodic memory, global-affect modulation, and top-down reconstruction while preserving equivalent capacity.

MEM predicts the formation of a representational core shared across modalities but flexibly linked to different regulatory contexts and action programs. The expected signal is reactivation of a specific ensemble during recall and its reinstatement in the appropriate lower-level map. RPT, PP, and GNWT need not deny such phenomena, but they do not make the semblion architecture their central mechanism. The result would therefore support MEM's specific organization of memory, but would not itself be a sufficient indicator of phenomenality; similar binding may improve the behavior of an unconscious system.

### 8.5 Tests Across Biological and Artificial Systems

**Test 5 Local Perceptual Phenomenality vs. Global Access in Humans**

To distinguish MEM and RPT from GNWT/GWT, paradigms that limit reporting should be used: binocular rivalry, bistable perception, task-irrelevant stimuli, and delayed reports. Changes in perceptual content are estimated in parallel through involuntary oculomotor or autonomic indicators and occasional control trials. The principal question is whether stable, content-specific recurrence in sensory regions and lower-level receptor maps also occurs in the absence of current reports and extensive frontoparietal activity. Activity must be temporally separated into stimulus processing, perceptual switching, decision, and response execution.

RPT would be supported if perceptual content and changes in it were explained by local sensory recurrence without the need for late global ignition. GNWT would be supported if conscious content consistently required nonlinear amplification and broad broadcasting, including under no-report conditions. MEM predicts the possibility of basic perceptual feeling with receptor-grounded re-entry and affective modulation before the content becomes available for elaborate reporting. It does not, however, predict that global access is always absent or irrelevant; global access may be a later capacity that uses already formed content.

A finding of ‘local re-entry without a frontal marker’ would not definitively exclude GNWT because the absence of a detected signal may reflect measurement sensitivity, and a no-report indicator may misclassify the percept. Replication using multiple methods, power analysis for null results, and an intervention test are therefore required. For MEM, a particularly informative case would be one in which perturbing receptor reconstruction changes an involuntary measure of content and a later preference, whereas perturbing reporting or the global workspace disrupts the declaration without producing an analogous change in the two earlier measures.

**Test 6. Cephalopods**

Cephalopods are a useful comparative case because they show learning, memory, and flexible behavior despite having a nervous system very different from that of mammals. Their distributed neural organization and the absence of linguistic reports allow us to ask whether perceptual and affective processes can occur without human-like language, narrative, or frontal cortex. Defensive behavior alone, however, is not evidence of pain or phenomenality.

The experimental program should combine mild, reversible nociceptive stimulation with measures of immediate response, persistent changes in motivation, conditioned place aversion or preference for pain relief, and flexible trade-offs between avoidance and other goals. Studies in octopuses have reported location-specific avoidance, preference for contexts associated with analgesia, wound-directed behavior, and neural responses to noxious stimulation (Crook, 2021). Similar work in cuttlefish has examined wound-directed behavior and its reduction following local analgesia (Kuo et al., 2022). These findings go beyond simple reflexes but still require controls for sedation, motor impairment, nonspecific drug effects, and learning unrelated to nociception.

A MEM-specific extension would examine whether three processes are coupled in the same animal: a receptor- or location-specific representation of the stimulus, reactivation of that representation after the stimulus has ceased, and modulation by internal regulatory state, memory, and available actions. Where feasible, neural recordings should be combined with behavioral measures. A reversible disruption of feedback could test whether persistent and context-dependent responses depend on recurrent processing while leaving basic reflexes and locomotion relatively intact.

Such a result would have contrastive significance. Recurrence alone would be compatible with RPT; threat prediction and prediction error could be explained by PP; and flexible motivational trade-offs could arise from homeostatic reinforcement learning. MEM would receive stronger support if receptor-grounded reactivation, regulatory-affective state, episodic memory, and flexible action were causally linked. The absence of such coupling would weaken the specific MEM interpretation but would not by itself establish the absence of phenomenal experience.

**Test 7. Simple Invertebrates**

The aim of this test is not to determine whether simple invertebrates are phenomenally conscious, but to examine whether they exhibit combinations of mechanisms that, according to MEM, may be relevant to phenomenality. The comparative program may include gastropods such as *Aplysia* and *Lymnaea*, insects such as bumblebees and *Drosophila*, and annelids such as *Eisenia* earthworms. *Caenorhabditis elegans* may serve as a lower-bound control for nervous-system complexity. The outcome should therefore be a graded profile of experimentally observed mechanisms and behaviors, not a ranking or binary diagnosis of consciousness.

For each taxon, observations should be separated into four layers: (a) receptors, nociceptors, and regulatory circuits; (b) learning, memory, protective behavior, and motivational trade-offs; (c) evidence for recurrent or content-specific reactivation; and (d) interpretation within competing theoretical frameworks, including MEM, RPT, PP, GNWT, and UAL. Behavioral evidence should not by itself be treated as evidence for phenomenality, while the absence of demonstrated recurrence should not be interpreted as evidence that recurrence is absent.

A common experimental protocol can examine responses to a mild, reversible aversive stimulus and to appropriate controls. Relevant measures include immediate withdrawal, persistent sensitization, protective behavior, avoidance or instrumental termination, context-dependent learning, preference for contexts associated with relief, trade-offs between avoidance and reward, and generalization to novel contexts. Existing studies provide examples of such phenomena in *Aplysia* (Walters, 1987), *Drosophila* (Khuong et al., 2019), earthworms (Wilson et al., 2014), and bumblebees (Gibbons et al., 2022). These findings demonstrate forms of nociception, learning, adaptation, or motivational flexibility; considered individually, none establishes subjective feeling.

A more specific MEM-related test would examine whether a previously experienced stimulus can be reactivated after the stimulus has ceased and whether this reactivation is influenced by internal regulatory state, memory, and available actions. Where technically feasible, behavioral measurements can be combined with electrophysiology, calcium imaging, or recordings from central and peripheral neural structures. The critical distinction is between content-specific reactivation and more general arousal, sensitization, or motor preparation.

Controls should exclude simpler explanations such as nonspecific arousal, altered locomotion, sensory impairment, hunger, fatigue, phototaxis, chemotaxis, or simple stimulus–response learning. Appropriate controls include sham or reward-only conditions, pseudoconditioning, novel contexts, yoked stimulation, and counterbalancing of stimulus location and modality. Key criteria and alternative explanations should be specified in advance and, where possible, replicated across laboratories.

The strongest MEM-relevant result would be a demonstrated coupling, within the same organism and time course, between three components: a receptor- or location-specific representation of the stimulus, its reactivation after the original input has ceased, and modulation of this reactivation by regulatory-affective state, memory, and action possibilities. Reversible disruption of feedback could provide a causal test of this coupling. Such a result would provide evidence for an organization that is more specifically consistent with MEM, while individual components would remain compatible with alternative explanations such as RPT, predictive processing, or homeostatic reinforcement learning.
Importantly, failure to demonstrate this coupling would not establish that the organism lacks phenomenal experience. It would instead limit the evidence for the specific MEM mechanisms under investigation. Thus, this test should be interpreted as comparative evidence about candidate organizational mechanisms, rather than as a direct test or measure of consciousness in simple invertebrates.

**Test 8 Synthetic MEM Agent**

In a robotic or simulated environment, a baseline agent, model-based RL, a recurrent world model, homeostatic RL, active inference, and a complete MEM-like agent perform identical task sequences under varying energy use, temperature, sensor integrity, and risk of damage. All variants receive the same budgets for data, parameters, memory, and computation. The tasks should require choices between immediate reward and long-term stability, preventive action before failure, transfer of a rule to a new map, and imaginal evaluation of action outcomes before execution. Measures include task success, cumulative regulatory cost, number of failures, accuracy of sensory reconstruction, prediction calibration, and transfer.

A full factorial ablation design comprising eight architectural variants will be used . The eight variants comprise the complete MEM-like architecture, three variants in which one function is disabled (regulatory interoceptive coupling, global affective modulation, or re-entry), three variants in which two functions are disabled, and one variant in which all three are disabled. The roles of episodic memory and semblions are assessed in additional, separate ablation experiments. In each variant, the remaining architectural functions should remain unchanged, and differences in parameter and operation counts should be controlled by functionally nonspecific layers or additional memory capacity.

MEM predicts not only an advantage for the complete architecture but also interactions among the functions under study, that is, dependence of the effect of one function on the presence of the others. Global affective modulation should influence action selection and representational consolidation, especially when it uses interoceptive information, while re-entry should enable predicted consequences to be reinstated in the appropriate sensory or interoceptive maps. If a resource-matched recurrent world model or active-inference system achieves the same outcome profile without the coupling of these functions postulated by MEM, the result will not be specific to MEM.

### 8.6 Auxiliary Prediction: Procedural Gaps

**Test 9 Procedural Gaps**

The agent is given a repertoire of known operators and a task requiring a novel combination of them. LLM planning, model-based RL, a recurrent world model, and a MEM-like system are compared. The procedural-gap mechanism detects the absence of a program exceeding a specified value threshold, generates candidates, simulates their consequences, selects the program with the lowest predicted regulatory cost, and consolidates the solution. Measures include the number of trials, cost, effectiveness, transfer to altered conditions, and stability after a delay. Ablations separately remove gap detection, simulation, global affect, or consolidation, while a capacity control preserves equivalent resources.

This test assesses creative planning and the integration of memory with regulation, but bears less directly on phenomenality than tests of receptor re-entry and affect. A positive result would support MEM's cognitive extension and the claim that the formalism is novel in its integration; by itself, it would not distinguish a feeling from a non-feeling system. It should therefore be designated in the preregistration as an auxiliary prediction rather than a critical criterion of consciousness.

### 8.7. Joint Support Criterion, Competing Explanations, and Falsification

Before research begins, a matrix of divergent predictions should be constructed. For each test, the theories should specify the required substrate or region, the predicted direction of information flow, the temporal window, dependence on report, dependence on regulatory state, and the effect of selective perturbation. The analysis should then compare not only the accuracy of individual predictions but the joint credibility of the models on new data. A hybrid model, for example RPT combined with active inference and episodic memory, must be admitted as a genuine competitor; otherwise, MEM's broader scope would result from defining the theories unequally.

Strong support for MEM would require convergence of at least four critical results: the content of secondary perception is reconstructed in receptor-grounded maps; the reconstruction is causally necessary for vividness or flexible use of content; the system's internal state modifies this reconstruction, memory, and selection through global affect; and these relationships recur in perceptual and affective tasks and in embodied action. Semblions, episodic memory, and procedural-gap handling provide an additional but weaker layer of support. The most persuasive evidence would be convergence across humans, cephalopods, simpler invertebrates lacking linguistic reports, and synthetic systems, using different measurement methods but the same logic of divergent predictions.

MEM would be substantially weakened if: (a) secondary perception and its vividness remained intact after content-specific re-entry into lower-level maps had been effectively abolished; (b) decoupling regulatory interoception from global affect did not alter learning, memory, or selection; (c) component effects were purely additive and fully explained by additional capacity; or (d) a resource-matched RPT, PP, GNWT, or hybrid model predicted the data equally well without mechanisms specific to MEM. Failure of the procedural-gap test would weaken the cognitive component of the architecture, but would not suffice to reject its hypothesis concerning basic perceptual-affective phenomenality.

In a limited comparison with individual implementations of RPT, PP, and GNWT, the complete MEM architecture encompasses a broader profile of prespecified indicators: sensory recurrence, receptor grounding, regulatory interoception, affective modulation, associative memory, and coupling of content to action. This is an advantage in integrative scope, not proof of phenomenality or a demonstration of superiority over every possible system. Only the following weaker claim can be empirically justified: with resources matched, MEM explains more preregistered, mutually dependent indicators than the alternative implementations compared, and does so by means of a single coherent architectural coupling.

Such a program shifts the burden of empirical justification from system declarations and a single test to divergent predictions, causal interventions, and the accumulation of evidence. A result consistent with the full MEM profile would make attribution of a capacity for experience more rational and would trigger more stringent ethical safeguards. It would still neither resolve the problem of other minds nor establish that the postulated conditions are sufficient for phenomenality.

## 9. MEM, Identity Theory, and Multiple Realizability

Multiple realizability is a central assumption in cognitive science. If consciousness depends on functional organization rather than on a single specific biological substance, then certain forms of consciousness may also be implemented in artificial systems. This thesis is consistent with the functionalist tradition in philosophy of mind, but MEM gives it a more concrete form. The biophysical identity theory within MEM concerns biological consciousness: an experience is identical to a specific dynamic biophysical configuration of a neuronal-astrocytic network, not merely to an abstract function of that network (Galus, 2023b; Galus & Starzyk, 2020). It does not follow logically from this position that a functionally similar electronic configuration will produce phenomenality.

MEM therefore does not adopt substrate independence as an established assumption. It admits it only as an additional open hypothesis: an artificial substrate might prove capable of implementing causally relevant dynamics, but functional similarity would provide an indicator, not proof of identity or experience. Testing this hypothesis requires comparison of temporal organization, causal integration, memory, and regulatory loops, rather than input-output correspondence alone. Technical interoceptors must be self-monitoring channels coupled to regulation; a telemetry reading of energy or temperature that does not affect priorities, learning, or protection of integrity does not satisfy this definition.

Biology suggests that perception, affect, and imagination serve the regulation of organismic action. A system lacking its own costs and regulatory couplings satisfies fewer MEM indicators, even if it reports states exceptionally well. This does not justify the categorical claim that every other architecture is merely an imitation. An artificial agent need not possess human morphology or human sensory modalities. Unusual sensors might create representations inaccessible to humans. Even complete implementation of the predicted couplings would not allow us to determine in advance whether qualia arise or what character they would have.

An indicator-properties approach enables graded empirical justification without deciding subjectivity on the basis of a single test (Butlin et al., 2023). MEM proposes an additional hypothetical profile of indicators: persistent receptor maps, regulatory self-monitoring, affective modulation, semblions, re-entry, secondary perception, episodic memory, and procedural-gap handling. The joint and causally integrated presence of these properties would increase the rationality of further research and of cautiously attributing a capacity for experience to the system, without settling phenomenality. Because subjective states are not directly accessible from a third-person perspective, empirical justification should rest on preregistered tests of divergent predictions, causal interventions, and convergence of independent indicators. Such tests can strengthen or weaken MEM, but they do not prove the occurrence of phenomenal feelings.

## 10. Discussion and Conclusions

The analysis developed in this article supports a distinction between the evolutionary accumulation of cognitive capacities and the emergence of phenomenality. Biological evolution did not produce a simple sequence of stages of consciousness. Rather, it progressively integrated environmental coupling, regulation, embodiment, learning, memory, valuation, multimodal processing, and recurrent interaction between internal states and perceptual representations. Excitability and regulation preceded increasingly complex plastic networks; learning and memory became integrated with valuation and action; and, in more complex organisms, perception, internal state, memory, and action became tightly coupled. The relevant conclusion is therefore not that any particular biological structure constitutes consciousness, but that phenomenality becomes a more concrete scientific question when multiple organizational processes are causally integrated within the same embodied system.

This provides the first link in the argument developed here: evolutionary organization identifies candidate properties; MEM provides a framework for integrating them. Within MEM, receptor grounding, interoceptive regulation, affective significance, semblion-based integration, re-entry, secondary perception, and action-related memory are not independent modules that individually generate consciousness. They are components of a dynamically coupled architecture in which perception, internal state, motivation, memory, and action continuously influence one another.

### What does MEM add beyond advanced information processing?

Contemporary AI systems demonstrate that sophisticated information processing does not by itself imply the organization proposed by MEM. LLMs provide powerful language processing, reasoning, and reporting capabilities; hybrid systems add persistent memory, external models, and tools; VLA systems connect multimodal information with action; and embodied robots introduce continuous interaction with the physical environment. Yet the presence of these capabilities, individually or in combination, does not establish phenomenality.

MEM adds a different organizational requirement. Its central claim is that cognition relevant to phenomenality must be grounded in a persistent perception–regulation–action loop and dynamically coupled to the internal state of the organism or agent. A candidate machine phenomenal state is therefore not defined by fluent report, by a particular computational module, or by the mere integration of multiple information streams. It is defined as a dynamic whole-system state in which sensor-grounded modal content is recurrently stabilized and causally coupled to a suprathreshold representation of the system's current regulatory state (Galus, 2026b).

This distinction is important because it changes the explanatory target. The question is not simply whether a system performs functions that resemble those of a conscious organism, but whether its information processing is organized around persistent self-regulation, grounded perception, internal significance, recurrent reconstruction, and action consequences. These requirements provide measurable architectural targets rather than an a priori attribution of consciousness.

**What can actually be tested?**

The third link in the argument is the translation of the MEM proposal into architectural criteria and experimentally testable predictions. The analysis presented in this article identifies several such targets: receptor grounding, recurrent re-entry, interoceptive thresholds, affective modulation, causal coupling between perceptual and regulatory states, semblion formation, episodic binding, secondary perception, and the temporal integration of these processes.

The proposed tests are intended to distinguish these mechanisms from simpler explanations based on local information processing, behavioral performance, or verbal report. Human experiments can examine receptor-grounded re-entry, secondary perception, imagery, interoceptive modulation, and the distinction between local perceptual phenomenality and global cognitive access. Comparative animal studies can examine whether related organizational signatures occur across organisms with different nervous-system architectures. Artificial systems provide a further opportunity: a synthetic MEM agent can be constructed with controlled manipulation of allostatic variables, interoception, re-entry, secondary perception, affective modulation, and integrated memory. Its behavior and internal dynamics can then be compared with matched systems in which selected components or couplings are removed.

The mathematical MEM formulation provides an additional step toward testability by translating the proposed organization into variables, couplings, thresholds, selection rules, and predicted costs that can be implemented, simulated, and subjected to ablation experiments (Galus & Starzyk, 2026). The important methodological principle is that the same architectural criteria should be tested against competing explanations. A result should provide stronger support for MEM only when the predicted effect depends on the proposed organization rather than merely on increased computational complexity, additional memory, richer sensory input, or greater behavioral capability.
Thus, the scientific argument developed here is not simply that MEM can describe a possible conscious machine. It is that MEM converts an evolutionary and conceptual hypothesis into architectural criteria and then into experimentally testable predictions. The resulting framework permits strong inference through implementation, intervention, ablation, comparison, and attempted falsification.

**What remains unresolved?**

Several fundamental questions remain open. First, it has not been established that receptor-grounded re-entry and suprathreshold interoceptive coupling are necessary conditions of phenomenality. They are candidate conditions whose empirical status must be determined by comparative experiments. Second, even if the proposed mechanisms prove sufficient for the characteristic behavioral and neural signatures predicted by MEM, this would not by itself establish that an artificial system has subjective experience. The relation between measurable organizational states and phenomenality remains an unresolved philosophical and scientific problem.

A further unresolved issue concerns physical realization. MEM is formulated at the level of organization rather than a particular biological material. This makes the framework compatible with the possibility that similar organizational structures could be implemented in different physical substrates, but it does not establish that phenomenal states are multiply realizable. Whether the relevant organizational identity can be realized outside neuronal systems remains an empirical and philosophical question.

The possibility of artificial phenomenality also introduces an ethical issue that cannot be ignored. If an artificial system were deliberately constructed with persistent negative regulatory states, and if those states were accompanied by converging indicators of the kind proposed by MEM, the possibility of artificial suffering could not simply be dismissed because certainty about subjective experience is unavailable (Butlin et al., 2023; Long et al., 2024). For this reason, experimental development should initially favor reversible simulation, minimal and time-limited negative regulatory signals, non-self-amplifying dynamics, explicit stopping thresholds, state monitoring, and safe extinction procedures. Precaution should increase as the number and integration of relevant indicators increase.

Overall, the article proposes a continuous scientific chain *from* ***evolutionary organization to MEM, from MEM to architectural criteria, from criteria to implementation, from implementation to predictions, and from predictions to falsification***. This chain does not establish that consciousness has already been reproduced artificially. Its purpose is more precise: to provide a mechanistic and experimentally accessible framework for determining which forms of organization distinguish advanced information processing from the candidate organizational conditions of phenomenality.

MEM should therefore be regarded as a falsifiable architectural research program, rather than as a declaration that a particular artificial system is conscious. Its scientific value ultimately depends on whether the proposed organizational criteria generate reproducible predictions, survive interventions and ablations, outperform simpler competing explanations, and clarify the relationship between biological and artificial forms of cognition. The unresolved question is not merely whether machines can reproduce increasingly sophisticated behavior, but whether a sufficiently integrated organization of perception, regulation, memory, affect, motivation, and action can instantiate the conditions under which phenomenality could arise.